\documentclass[pra,aps,twocolumn,superscriptaddress]{revtex4-2}  

\usepackage{amsmath}  
\usepackage{amsfonts} 
\usepackage{graphicx} 
\usepackage{amssymb}
\usepackage{pstcol,times,xcolor,graphicx,graphics,pstricks,pst-node,pst-coil,pst-grad}
\usepackage[normalem]{ulem}
\usepackage{dsfont}
\usepackage{bm}					
\usepackage{hyperref}			
\usepackage{float}

\newcommand{\angstrom}{\text{\normalfont\AA}}

\graphicspath{{Figuras/}}

\begin{document}


\title{Controlling the atom-sphere interaction with an external electric field}

\author{P. P. Abrantes}
\email{ppabrantes91@gmail.com}
\affiliation{Instituto de F\'{\i}sica, Universidade Federal do Rio de Janeiro, 21941-972, Rio de Janeiro, Brazil} 

\author{V. Pessanha}
\email{victorpessanha15@gmail.com}
\affiliation{Instituto de Matem\'{a}tica, Universidade Federal do Rio de Janeiro, 21941-909, Rio de Janeiro, Brazil} 

\author{Reinaldo de Melo e Souza}
\email{reinaldos@id.uff.br}
\affiliation{Instituto de F\'{\i}sica, Universidade Federal Fluminense, 24210-346, Rio de Janeiro, Brazil}	

\author{C. Farina}
\email{farina@if.ufrj.br}
\affiliation{Instituto de F\'{\i}sica, Universidade Federal do Rio de Janeiro, 21941-972, Rio de Janeiro, Brazil} 




\begin{abstract}

We investigate the system constituted by a polarizable atom near a nanosphere under the influence of an external electrostatic field, showing that the attractive dispersive force between them can be overcome by the electrostatic interaction. Therefore, in addition to the advantageous possibility of actively tuning the resultant force with an external agent without the requirement of physical contact, this force may also become repulsive. We analyze this situation in different physical regimes of distance and explore the interaction of different atoms with both metallic and dielectric spheres, discussing which cases are easier to control. Furthermore, our results reveal that these repulsive forces can be achieved with feasible field intensities in the laboratory.

\end{abstract}


\maketitle


\section{Introduction \label{SecIntrod}}


Quantum electrodynamics (QED) describes light-matter interactions with unprecedented success, presenting an excellent agreement between theoretical predictions and experimental results. One of the many situations in which QED plays a key role is in the understanding of electromagnetic interactions between neutral bodies with no permanent multipoles, the so-called dispersive forces \cite{MilonniBook}, that arise from quantum fluctuations in their charge and current distributions. It is a multidisciplinary field that has sparked much interest for decades in different areas, such as chemistry, biology, colloid science, quantum field theory, and material science (see Refs. \cite{IsraelaBook, BuhmannBook} and references therein). These ubiquitous forces typically present an attractive character, a consequence of the fact that a fluctuating dipole in a given body induces a dipole in the bodies in its vicinity, which, in most cases, favors attraction, although this is not always the case. However, when the interaction is attractive, this may lead to undesirable effects in nano and micromechanical systems, such as adhesion and stiction \cite{Serry1998, buks2001, Broer2013, Sedighi2015, Rosa2018}. It has been an important issue and gave rise to a significant search for engineering configurations that exhibit repulsive forces. For a detailed overview of theoretical and experimental efforts regarding dispersive forces, see Refs. \cite{Galina2009, Woods2016, Galina2020, Gong2021, Laliotis2021}.

Unfortunately, we have very few general recipes that enable us to architect systems endowed with repulsive Casimir forces, and most situations require a complete calculation before we conclude their sign. Some examples of configurations in which repulsion can already be ruled out beforehand were stated by Kenneth and Klich \cite{Kenneth2006}. The authors proved that the interaction of (nonmagnetic) dielectric bodies or conductors is always attractive whenever they constitute a mirror-symmetric setup, independently of the objects' shape or their local dielectric functions. Interestingly, it was recently shown that a strategy to circumvent this restriction and achieve repulsion can be accomplished with the insertion of an intermediate chiral medium between the two materials \cite{Jiang2019}.

Actually, it was discovered long ago that repulsive Casimir force may be reached. In 1961, Dzyaloshinskii, Lifshitz, and Pitaevskii \cite{DLP1961} investigated systems constituted of three different nonmagnetic media, namely, two parallel semi-infinite homogeneous dielectrics of permittivities $\varepsilon_1 (\omega)$ and $\varepsilon_2 (\omega)$ separated by an infinite slab of a third dispersive homogeneous medium of permittivity $\varepsilon_3 (\omega)$. If the relation $\varepsilon_1~(i\xi)~<~\varepsilon_3~(i\xi)~<~\varepsilon_2~(i\xi)$ holds for a wide range of frequencies, repulsion occurs according to their theory. Experimentally, it had not been observed until 1996, when measurements were performed with an atomic force microscope in the van der Waals limit \cite{Milling1996}. In this same regime, subsequent evidence of repulsive interaction was also presented by other groups \cite{Meurk1997, Lee2001, Lee2002, Feiler2008}. In 2009, a direct measurement of long-range repulsive forces between a gold-covered sphere and a large silica plate mediated by bromobenzene was reported \cite{Munday2009}.

Another possible way of achieving repulsion concerns dielectric-magnetic materials. As shown by Feinberg and Sucher \cite{Feinberg1968, Feinberg1970}, this is exactly the case for two atoms if one of them is electrically polarizable while the other is magnetically polarizable. Based on these works, but in the context of stochastic QED, Boyer \cite{Boyer1974} verified repulsive forces when two parallel plates are placed close to each other with vacuum in between, if one of them is perfectly conducting (electric permittivity $\varepsilon \rightarrow \infty$) and the other one is perfectly permeable (magnetic permeability $\mu \rightarrow \infty$). Results involving dielectric-magnetic materials have also been discussed \cite{BuhmannBook, Farina2002a, Farina2002b}. In recent years, the search for repulsive forces has been expanded to topological materials \cite{Grushin2011a, Grushin2011b, Nie2013, RodriguezLopez2014, Wilson2015, RodriguezLopez2017, Lu2018} and metamaterials. In the latter case, attempts were carried out \cite{Yannopapas2009, Silveirinha2010, McCauley2010}, but it turned out that repulsion in such an assembly is exceedingly difficult \cite{Rosa2008a, Rosa2008b}. Likewise, systems out-of-thermal equilibrium have received special attention. Non-equilibrium Casimir forces arise in cases where the objects and the environment are kept at different temperatures, and some situations may be a possible source of repulsive interactions \cite{Gong2021, Messina2011, Bimonte2011, Iizuka12021}.

Some predictions have demonstrated that bodies with nontrivial geometry can present a great opportunity to generate repulsive Casimir interaction in vacuum. Interesting examples consist of a polarizable particle centered above an infinitely conducting \cite{Levin2010, McCauley2011} or dielectric \cite{Shajesh2012} plate with a circular hole. Particularly, in the case of the conducting plate, this repulsive nonretarded force was also analytically investigated  \cite{Eberlein2011, Reinaldo2011}. In addition, for sufficient anisotropy, repulsion can be obtained between an atom and some conducting bodies as, for instance, a semi-infinite plate, a wedge \cite{Milton2011}, and a cylinder, provided that the atom moves on a trajectory nonintersecting with the cylinder \cite{Milton2012}. Recently, the nonretarded force between a polarizable particle and a grounded conducting toroid has been investigated and repulsion was predicted depending on the values of its two radii and the distance from the particle to the geometrical center of the toroid \cite{Abrantes2018}.

However, if one has to deal with specific geometries and materials, the aforementioned routes will not help. To obtain repulsion in this scenario, we must tailor light-matter interactions at the nanoscale. The idea of tuning QED effects can be traced back to the pioneering work of Purcell in 1946 \cite{Purcell1946}, where it was shown that the environment of a quantum emitter may substantially affect its spontaneous emission rate. Controlling the dispersive interaction resorting to the application of tunable external electric or magnetic fields has proved to be a profitable venue \cite{Jiang2019, RodriguezLopez2017, Fuchs2018, Haug2019}. It was recently shown that an electrostatic field could generate a repulsive interaction between atoms \cite{Fiscelli2020}. The interpretation of their result was disputed \cite{Abrantes2021, Reply2021}, and subsequently, a thorough and rigorous study was carried out to settle the issue \cite{Tkatchenko1, Tkatchenko2}. However, regardless of the interpretation, Ref. \cite{Fiscelli2020} has correctly and interestingly shown that realistic values of electrostatic fields can be used to tune the interaction between atoms, changing it from attractive to repulsive.

A natural question is whether this holds for interactions between atoms and surfaces. In this regard, here we investigate the effect of applying an external electrostatic field in the system composed of a neutral and isolated sphere and an atom in its ground state with no permanent dipole moment. We analyze both metallic and dielectric spheres and  show that, by varying the intensity and the orientation of the electric field, the component of the resultant force on the atom along the line containing the center of the sphere and the atom, given by the sum of dispersive and electrostatic contributions, can become repulsive, meaning that the electrostatic force may overcome the dispersive one. Our findings reveal that such a control can be performed using the electric field as an effective knob that allows for the manipulation of the attractive or repulsive character of the atom-sphere interaction, requiring no physical contact. Most importantly, this tunability can be obtained even for feasible values of the external field, and can easily be implemented in laboratories. Furthermore, we show that our results are robust against the size of the sphere, enabling the same electrostatic field to control a system composed of different bodies interacting with the atom.

This paper is organized as follows. In Sec. \ref{AtomSphereInt} we introduce the relevant forces that take place in the setup and establish the validity conditions of our calculations. Section \ref{RD} is dedicated to the discussions of our results and Sec. \ref{Conclusions} is left for final remarks and conclusions. Additionally, we included an Appendix containing important information on the mathematical description of materials and atoms studied in this paper.


\section{Atom-sphere interaction}
\label{AtomSphereInt}

\begin{figure}[t!]
\begin{center}
\includegraphics[width=7.8cm]{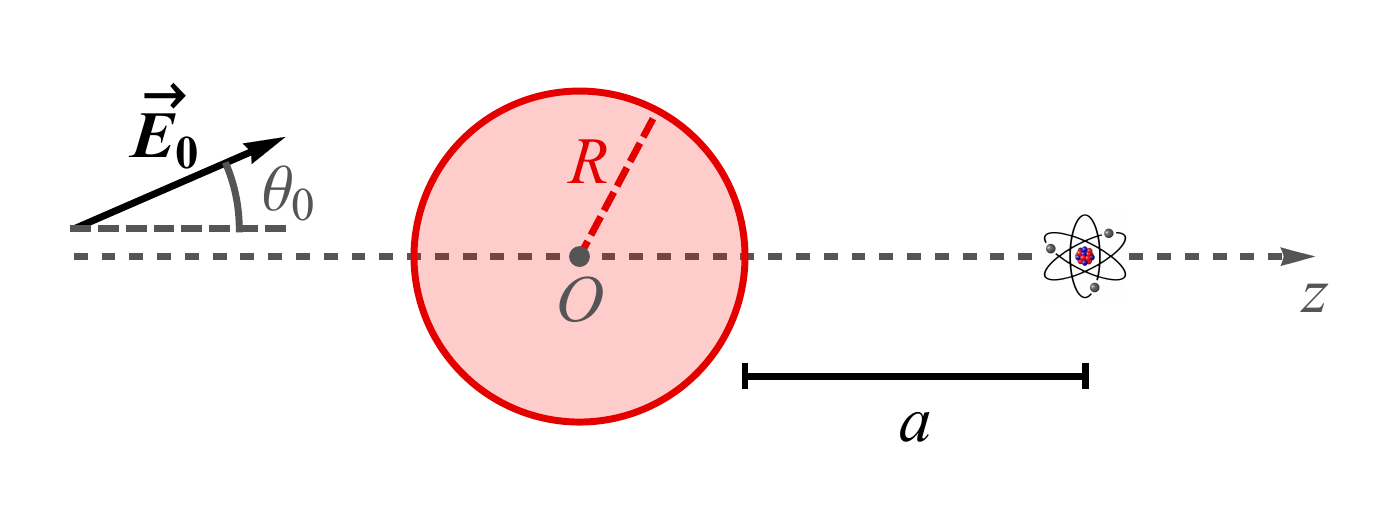}
\end{center}
\vskip -0.6cm
\caption{A neutral and isolated sphere of radius $R$ and a polarizable atom in its ground state at a distance $a$ from the surface of the sphere. The $Oz$ axis is chosen parallel to the line connecting the atom to the center of the sphere. An external uniform electrostatic field $\bm{E}_0$ is applied at an angle $\theta_0$ with respect to the $Oz$ axis.}
\label{AtomSphere}

\end{figure}

In this paper, we are concerned with the physical system depicted in Fig. \ref{AtomSphere}. It consists of a neutral and isolated sphere of radius $R$ and a polarizable atom in the ground state with no permanent dipole moment (nor higher-order multipoles). For convenience, we choose the Cartesian axes in such a way that the sphere is centered at the origin and the atom is located at a generic point $z_a = R + a$ of the positive semi-axis $Oz$, with $a$ being the distance from the atom to the surface of the sphere. This system is exposed to an external uniform electrostatic field $\bm{E}_0$ applied at an arbitrary angle $\theta_0$ with respect to the $Oz$ axis. The atom interacts with the sphere through a dispersive force and the applied electric field also induces an electrostatic coupling between them.

In the following subsections, we separately present the evaluation of each contribution to analyze which effect is dominant.


\subsection{Dispersive force}


Before presenting the expression of the dispersive force exerted by the sphere on the atom, a comment is in order. In principle, this dispersion force is modified by the external electrostatic field since both energy levels and eigenstates of an atom are affected by the electrostatic field. As a consequence, the atomic polarizability of the ground state of the atom is also altered. However, for the field intensities and range of distances between the atom and the sphere to be considered in this paper, it can be shown that such a variation is negligible. A heuristic way of realizing this is to consider the atom as a two-level oscillating system with frequency $\omega_0$ and reduced mass $\mu$. A constant force, such as the one exerted by an electrostatic uniform field, only shifts the equilibrium position, leaving the oscillations unaltered. Hence, the polarizability is unaffected as long as the harmonic approximation holds during the whole motion. This requires $e E_0/(\mu\omega_0^2)\ll a_0$, where $e$ is the electron charge, $E_0$ stands for the applied electrostatic field, and $a_0$ is the Bohr radius. Taking $a_0 \sim 0.5$ $\angstrom$, $\mu$ as the electron mass, and $\omega_0^2 \sim 10^{31}$ Hz$^2$ (see Appendix \ref{AppendA}), it is possible to confirm that we may neglect changes in the atomic polarizability for fields satisfying $E_0 \ll 10^{10}$ V/m, a condition met by the fields employed in this paper. A full quantum-mechanical calculation yields an equivalent conclusion. As shown in Ref. \cite{Abrantes2021}, the relative change in the static polarizability is given by the Stark effect, $\delta\omega_0/\omega_0 \sim \alpha E_0^2/(\hbar\omega_0)$, which shows that we may neglect variations in the polarizability for all cases in which perturbation theory remains valid. For cesium, the atom with the greatest polarizability considered here, the relation $\delta\omega_0/\omega_0\ll 1$ requires $E_0 \ll 10^{10}$ V/m. Hence, we may comfortably disregard any effect of $\bm{E}_0$ in the dispersive interaction, as can also be shown by a full quantum electrodynamics analysis \cite{Tkatchenko1, Tkatchenko2}.

To begin with, we consider the small sphere limit ($R \ll a$), so that the dipole approximation holds. In this case, assuming a non-magnetic sphere, the dispersive interaction energy is given by \cite{thiru, Hemmerich2016}
\begin{align}
	U^{\rm (disp)}_{\rm dip} (z_a) &= - \frac{\hbar R^3}{4 \pi^2 \varepsilon_0 z_a^6} \int_{0}^{\infty} \!\!\! d\xi \, \alpha_a(i \xi) \frac{\varepsilon(i \xi) - 1}{\varepsilon(i \xi) + 2} e^{-2\xi z_a/c} \nonumber \\
	&\times  \left[ 3 + \frac{6 \xi z_a}{c} + \frac{5 (\xi z_a)^2}{c^2} + \frac{2 (\xi z_a)^3}{c^3} + \frac{(\xi z_a)^4}{c^4} \right] \,,
\label{UCPRed}
\end{align}
where $\varepsilon_0$ is the electric permittivity of vacuum, $c$ is the light velocity in vacuum, $\alpha_a (i \xi)$ denotes the atomic polarizability evaluated at imaginary frequencies $\xi$, and we used that the sphere dynamical polarizability can be written as \cite{ZangwillBook}
\begin{equation}
	\alpha_s (i \xi) = 4 \pi \varepsilon_0 \frac{\varepsilon(i \xi) - 1}{\varepsilon(i \xi) + 2} R^3\, , \label{alphas}
\end{equation}
where $\varepsilon(i \xi)$ is its electric permittivity (normalized by $\varepsilon_0$). In the next section, we shall perform the integration in Eq. (\ref{UCPRed}) numerically for different materials and atoms. However, there are some limits for which we may obtain simple analytical expressions, useful for a first analysis, since they can already furnish a physical intuition about the orders of magnitude of all quantities involved. In the case of a two-level atom, we can employ a Lorentz oscillator model with a single resonance as indicated in Eq. (\ref{alpha}) of the Appendix. Furthermore, for a perfectly conducting sphere, Eq. (\ref{alphas}) simplifies to 
\begin{equation}
	\alpha_s^{\rm (c)} = 4\pi \varepsilon_0 R^3 \,.
\label{alphasc}
\end{equation}
Substituting these results back into Eq. (\ref{UCPRed}), we may arrive at closed formulas both for the retarded regime ($z_a \gg c/\omega_0$, where $\omega_0$ is the atomic transition frequency) and for the non-retarded one ($z_a \ll c/\omega_0$). In the retarded regime, we may take $\alpha_a (i \xi) \approx \alpha_a (0)$ in Eq. (\ref{UCPRed}) \cite{thiru}, with $\alpha_a (0)$ denoting the static atomic polarizability, and, by performing the integration, we obtain
\begin{equation}
	U^{\rm (disp)}_{\rm dip,R} = - \frac{23 \hbar c \alpha_s^{\rm (c)} \alpha_a (0)}{64 \pi^3 \varepsilon_0^2 z_a^7} \,.
\label{udr} 
\end{equation}
On the other hand, in the non-retarded regime, Eq. (\ref{UCPRed}) reduces to
\begin{align}
	U^{\rm (disp)}_{\rm dip,NR} &=  - \frac{3 \hbar}{16 \pi^3 \varepsilon_0^2 z^6} \int_{0}^{\infty} \!\! d\xi \, \alpha_s^{\rm (c)} \alpha_a (i \xi) \nonumber \\
	&= - \frac{3 \hbar \omega_0 \alpha_s^{\rm (c)} \alpha_a (0)}{32 \pi^2 \varepsilon_0^2 z_a^6} \, .
\label{udnr}
\end{align}
This expression is usually not very realistic, since the non-retarded and the perfectly conducting sphere approximations are not compatible in general. As a matter of fact, in this distance regime, the distance $a$ must be much smaller than the dominant transition wavelength $\lambda_0 = 2 \pi c/\omega_0$ of the atom such that retarded effects in the interaction can be disregarded. In addition, the perfect conductor approximation assumes that the electric field does not penetrate the material, holding, thereby, as long as $a$ is much greater than the penetration length $l_p$. Therefore, these two approximations together are valid for $l_p \ll a \ll \lambda_{0}$, which restricts their applicability regime. However, it will be useful as a first rough description to investigate orders of magnitude.

When $a \lesssim R$, the dipole approximation for the sphere breaks, implying that Eq. (\ref{UCPRed}) is no longer valid, and we must now sum over much more multipoles. Assuming the non-retarded regime, we may obtain the dispersive interaction from the expression
\begin{eqnarray}
    U^{\rm (disp)}_{\rm NR} &=& -\frac{\hbar}{8\pi^2\varepsilon_0} \sum_{l=1}^{\infty} (2l+1) (l+1) \cr\cr
     &\times& \frac{R^{2l+1}}{z_a^{2l+4}}\int_0^{\infty} \!\! d\xi \, \alpha_a (i \xi) \frac{\varepsilon(i \xi) - 1}{\varepsilon(i \xi) + [(l+1)/l]} \, , \label{buhmanneq}
\end{eqnarray}
as shown in Ref. \cite{Hemmerich2016}. Let us now analyze the electrostatic part of the interaction.


\subsection{Electrostatic force}


In this subsection, we describe the electrostatic force that arises between the atom and the sphere due to the application of an electrostatic field ${\bm E}_0$. In such a situation, the atom acquires an induced electric dipole moment in response to the total field acting on it, namely, the external field and the electrostatic field created by the sphere. By the same token, in principle, the sphere also suffers the influence of a resultant electric field given by the sum of the external field and the field created by the polarized atom. However, the atomic polarizability of an atom in its ground state scales with $4 \pi \varepsilon_0 a_0^3$, and the electric field intensity of a point dipole of magnitude $p$ behaves as $\sim p/(4\pi \varepsilon_0 r^3)$, with $r$ being the distance of the space point under consideration to the dipole. Hence, the ratio between the magnitude of the field created by the atom on the sphere and that of the external field is on the order of $\sim (a_0/a)^3$, a negligible quantity in our configuration. As a consequence, although the electric dipole moment induced on the atom will be computed taking into account the external field and the electrostatic field created by the sphere, it will be enough to consider only the external field $\bm{E}_0$ when computing the response of the sphere to the electrostatic field acting on it.

It is well-known that the electrostatic field created by the sphere when exposed to an external uniform and constant field ${\bm E}_0$ is identical, in its outer region, to that of a point dipole located at its center \cite{ZangwillBook}, which we shall denote by ${\bm p}_s =  \alpha_s (0) {\bm E}_0$, with $\alpha_s$ given in Eq. (\ref{alphas}). The electrostatic field created by the sphere at the atom's position is given by
\begin{equation}
	{\bm E}_s ({\bm r}_a) =  \frac{\alpha_s}{4 \pi \varepsilon_0}\frac{3 E_0 \,\mbox{cos}\theta_0 \,\bm{\hat{z}} - {\bm E}_0 }{z_a^3}\, ,
\label{eqn: E-Esfera}
\end{equation}
so that, the electric dipole moment induced on the atom by the resultant field, ${\bm p}_a = \alpha_a \left[{\bm E}_0 + {\bm E}_s ({\bm r}_a)\right]$, takes the form
\begin{equation}
	{\bm p}_a  = \alpha_a \left( 1 -  \frac{\alpha_s}{4 \pi \varepsilon_0 z_a^3}\right) {\bm E}_0 + \frac{3 \alpha_a \alpha_s E_0\,\mbox{cos}\theta_0}{4 \pi \varepsilon_0 z_a^3}\, {\bm{\hat z}} \,.
\label{eqn: pa}
\end{equation}
Hence, the electrostatic force exerted by the sphere on the atom is the same as that exerted by a point dipole ${\bm p}_s$ at the origin on a point dipole ${\bm p}_a$ at the atom's position. It is worth emphasizing that we are only interested in the $z$ component of this force. In this sense, we initially write the general expression for the force exerted by a dipole $\bm{p}'$ at the origin on a dipole $\bm{p}$ at an arbitrary position $\boldsymbol{r}$, given by
\begin{eqnarray}
	{\bm F}^{\rm (el)}_{{\bm p}{\bm p}^\prime} &=& \frac{1}{4 \pi \varepsilon_0 r^{4}}\Bigl[ 3 ({\bm p} \cdot \bm{\hat{r}) {\bm p}^\prime} \;+ \; 3 ({\bm p}^\prime \cdot \bm{\hat{r}}) {\bm p} \cr\cr
	&+& \; 3 ({\bm p} \cdot {\bm p}^\prime) \bm{\hat{r}} \; - \; 15 ({\bm p}\cdot \bm{\hat{r}}) ({\bm p}^\prime \cdot \bm{\hat{r}}) \bm{\hat{r}} \Bigr]\, .
\label{eqn: dip-dip-force}
\end{eqnarray}
The $z$ component of the electrostatic force $F_z^{\rm (el)}$ on the atom can be obtained by making the substitutions ${\bm p}^\prime \longrightarrow {\bm p}_{s}$, ${\bm p} \longrightarrow {\bm p}_{a}$, ${\bm{\hat r}} \longrightarrow {\bm{\hat z}}$, and taking the scalar product of both sides of Eq. \eqref{eqn: dip-dip-force} with $\bm{\hat{z}}$, which leads us to
\begin{equation}
	F^{\rm (el)}_z = \frac{3}{4 \pi \varepsilon_0 z_a^4}\Bigl[{\bm p}_a \cdot {\bm p}_s \; - 3 ({\bm p}_a \cdot \bm{\hat{z}}) ({\bm p}_s \cdot {\bm{\hat z}}) \Bigr] \,.
\label{eqn: Fz}
\end{equation}
The last two scalar products readily follow from Eq. \eqref{eqn: pa} and the fact that ${\bm p}_s = \alpha_s {\bm E}_0$, while the first one is given by 
\begin{equation}
	{\bm p}_a \cdot {\bm p}_s = \alpha_a \alpha_s E_0^2 \left[ 1 + \frac{\alpha_e}{4 \pi \varepsilon_0 z_a^3} (3\,\mbox{cos}^2\theta_0 - 1)\right] \, .
\label{eqn: pa.pe}
\end{equation}
After plugging these results back into Eq. \eqref{eqn: Fz}, we are left with
\begin{align}
	F_{z}^{\rm (el)} &= \frac{3 \alpha_{a} \alpha_{s} E_{0}^{2}}{4 \pi \varepsilon_{0} z_{a}^{4}} \nonumber \\
	&\times \left[ 1 - \frac{\alpha_{s}}{4 \pi \varepsilon_{0} z_{a}^{3}} - 3 \cos^{2}\theta_{0} \left( 1 + \frac{\alpha_{s}}{4 \pi \varepsilon_{0} z_{a}^{3}}\right)  \right] \,.
\label{eqn: Fz_res_final}
\end{align}
To check the consistency of this formula, let us consider two particular configurations. First, we choose $\theta_{0} = 0$ so that ${\bm p}_s$ and ${\bm p}_a$ are aligned and the force between them is attractive. Indeed, we see that Eq. $\eqref{eqn: Fz_res_final}$ becomes
\begin{equation}
	F_{z}^{\rm (el)} = -\frac{6 \alpha_{a} \alpha_{s} E_{0}^{2}}{4 \pi \varepsilon_{0} z_{a}^{4}}\left( 1 + 2 \frac{\alpha_{s}}{4 \pi \varepsilon_{0} z_{a}^{3}} \right) < 0\, .
\label{fzel0}
\end{equation}
If we now consider $\theta_{0} = \pi/2$, a situation in which the dipoles are parallel to each other but perpendicular to the $Oz$ axis, we obtain a repulsive force. In fact, it follows from Eq. (\ref{eqn: Fz_res_final}) that
\begin{equation}
	F_{z}^{\rm (el)} = \frac{3 \alpha_{a} \alpha_{s} E_{0}^{2}}{4 \pi \varepsilon_{0} z_{a}^{4}} \left(1 - \frac{\alpha_{s}}{4 \pi \varepsilon_{0} z_{a}^{3}} \right) \,,
\label{fzelpisobre2}
\end{equation}
which is clearly a positive expression since $\alpha_s/(4\pi\varepsilon_{0}z_{a}^{3}) < 1$.

As a final comment, we point out that $\alpha_s/(4 \pi \varepsilon_0 z_a^3) \sim (R/z_a)^3$, according to Eq. (\ref{alphas}) and, therefore, whenever we work in the small sphere limit ($R \ll a$), we may simply approximate Eq. (\ref{eqn: Fz_res_final}) by
\begin{equation}
	F_{z}^{\rm (el)} \approx \frac{3 \alpha_{a} \alpha_{s} E_{0}^{2}}{4 \pi \varepsilon_{0} z_{a}^{4}} \left(1 - 3\cos^{2} \theta_{0} \right) \, . 
\label{fzeldip} 
\end{equation}


\section{Results and discussions}
\label{RD}

We start by considering the system composed of a hydrogen atom near a gold sphere, demonstrating that an external electrostatic field can be applied to control the attractive or repulsive character of the resultant force. Next, we show that the same behavior also occurs when considering other atomic species, such as Na, K, Rb, Cs, and Fe. Finally, to attest that the tuning mechanism we are discussing here is robust under the change in the sphere's material, we analyze the case of a dielectric sphere made of silicon dioxide (SiO$_2$). In this situation, we verify that repulsive resultant forces can also be obtained within the same approach with the only difference that, for a given distance between the atom and the sphere, the values of necessary fields for switching the attractive or repulsive character of the force are slightly different.


\subsection{Hydrogen atom near a metallic sphere \label{SubSecA}}


In this subsection, we discuss the interaction between a hydrogen atom and a gold sphere. The model for the atomic polarizability and the dielectric function of the sphere can be found in the Appendix. To facilitate the analysis of the change in sign of the resultant force, we define the ratio $\Gamma$ between the $z$ component of the resultant force and the absolute value of the dispersive force acting on the atom, namely,
\begin{equation}
	\Gamma = \frac{F_z^{\rm (net)}}{\vert F^{\rm (disp)}\vert} = \frac{F_z^{\rm (el)} + F^{\rm (disp)}}{\vert F^{\rm (disp)}\vert}  \,,
\label{gamma}
\end{equation}
where $F^{\rm(disp)} = - \partial U^{\rm (disp)}/\partial z_a$, with $U^{\rm (disp)}$ being the dispersive interaction energy. According to the previous definition, $\Gamma > 0$ means a repulsive force on the atom, while $\Gamma < 0$ means an attractive one. Throughout this paper, we will mainly focus on discussions of the ratio $\Gamma$ calculated with the dispersive interaction given in Eq. (\ref{UCPRed}), which is valid for any distance $z_a$ as long as obeying the dipole approximation. In a few situations, however, we will also show results for $\Gamma$ considering distances $a$ within the range of the non-retarded regime. In these cases, we shall resort to Eq. (\ref{buhmanneq}) to evaluate the dispersive contribution. Nonetheless, we will call the reader's attention every time this change occurs to avoid any misinterpretation.

\begin{figure}[t!]
	\centering
	\includegraphics[scale =0.55]{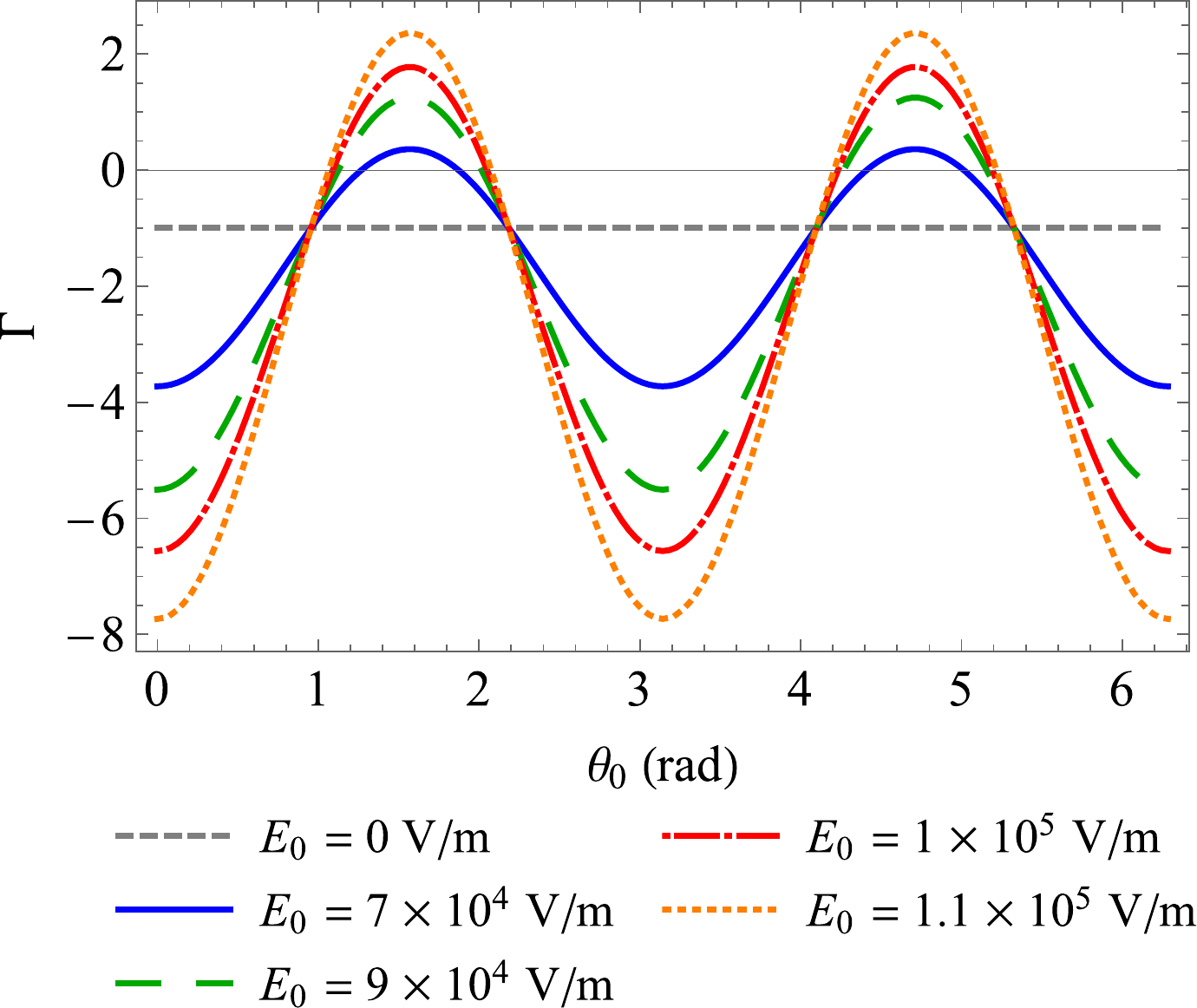}
	\caption{Ratio $\Gamma$ as a function of the angle $\theta_0$ between the applied electric field and the $Oz$-axis. Different colors refer to different field intensities. We set $R = 60$~nm and $a = 700$~nm.}
	\label{fig: PltHvarth700}
\end{figure}

We begin assuming that the distance separating the atom and the sphere is much larger than the sphere's radius ($a \gg R$), and therefore we employ Eq. (\ref{UCPRed}) in order to evaluate the dispersive force. Note that, in doing so, some care must be taken in choosing these parameters. To determine the direction of application of the field that provides the strongest repulsion, Fig. \ref{fig: PltHvarth700} displays $\Gamma$ as a function of $\theta_0$ for feasible intensities $E_0$ of the electrostatic field. We set the sphere's radius $R = 60$~nm and the distance between the atom and the sphere $a = 700$~nm, typical values in experiments involving dispersive forces. First of all, a direct inspection of Fig. \ref{fig: PltHvarth700} reveals that there are intervals of $\theta_0$ for which $\Gamma$ is positive, which implies a repulsive force between the sphere and the atom. It is also evident that the orientations of the electrostatic field that best favor repulsion occur for $\theta_0 = \pi/2$ or $\theta_0 = 3\pi/2$ as expected, since the induced dipoles on the atom and the sphere are perpendicular to the $Oz$ axis and parallel to each other in both cases, being the configurations of maximal electrostatic repulsion. Another interesting trait present in Fig. \ref{fig: PltHvarth700} is that the absolute value of $\Gamma$ at the peaks is smaller than its absolute value at the valleys. This follows from the combination of two intertwined features: {\it (i)} the repulsive electrostatic force for $\theta_0 = \pi/2$ is less intense than the attractive electrostatic force for $\theta_0 = 0$, as can be seen from Eqs. (\ref{fzel0}) and (\ref{fzelpisobre2}), and {\it (ii)} the electrostatic and dispersive contributions to the resultant force have opposite directions for $\theta_0 = \pi/2$, whereas they point in the same direction for $\theta_0 = 0$. Finally, one should note that the intersection of all curves occurs at $\Gamma = - 1$, a direct consequence of the definition in Eq. (\ref{gamma}). $\Gamma = -1$ implies $F^{\rm (el)}_z = 0$ and, since $F_z^{\rm (el)}$ is proportional to $E_0^2$ [see Eq. (\ref{eqn: Fz_res_final})], the solution of this condition is independent of $E_0$.

\begin{figure}[t!]
	\centering
	\includegraphics[scale =0.55]{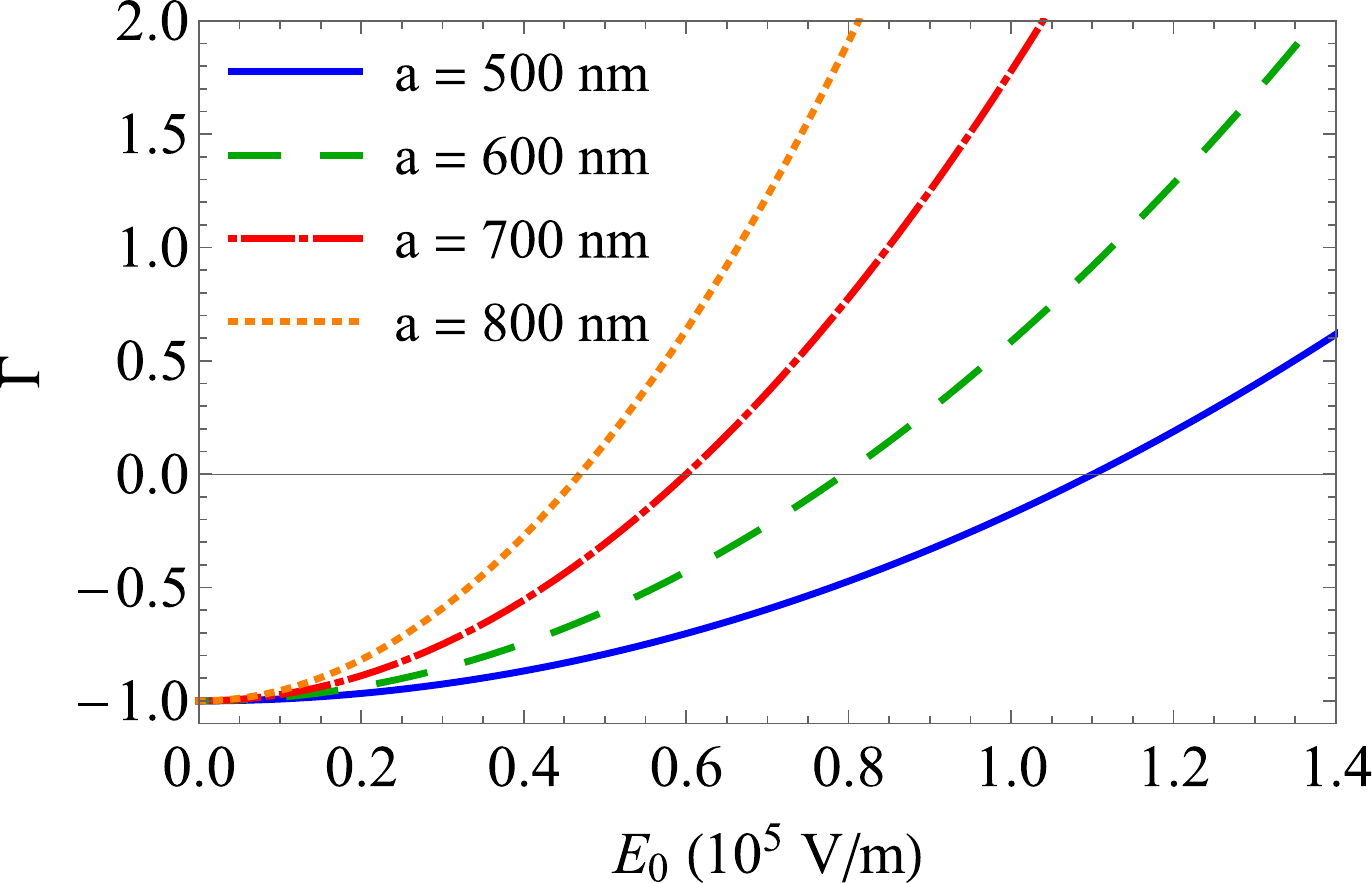}
	\caption{Ratio $\Gamma$ as a function of the field intensity $E_{0}$. Different colors denote different distances $a$ from the atom to the sphere's surface. We set $R = 60$~nm and $\theta_0 = \pi/2$.}
	\label{fig: PltHvarE0}
\end{figure}

\begin{figure}[b!]
	\centering
	\includegraphics[scale =0.55]{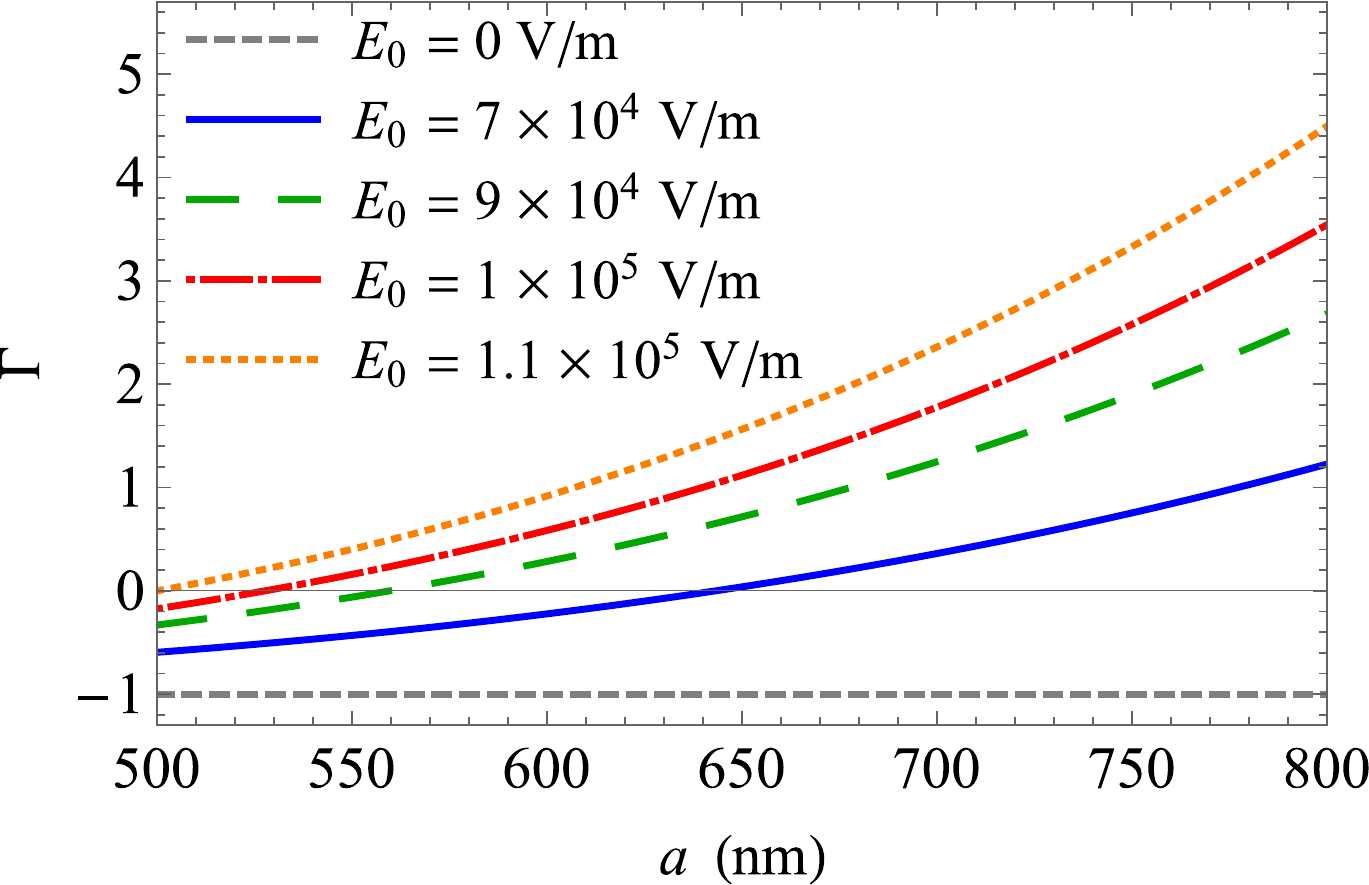}
	\caption{Ratio $\Gamma$ as a function of the distance $a$ between the atom and the sphere's surface for different field intensities. We set $R = 60$~nm and $\theta_0 = \pi/2$.}
	\label{fig: PltHvara}
\end{figure}

Let us now investigate the dependence of $\Gamma$ with the continuous variation of the other parameters, keeping $\theta_0$ fixed at $\pi/2$. In Fig. \ref{fig: PltHvarE0}, we plot $\Gamma$ as a function of $E_0$ for different values of the distance $a$ and $R = 60$~nm. For a given value of $a$, as we increase the magnitude of the external field $E_0$, the system undergoes from an attractive to a repulsive resultant force. Moreover, this can be achieved with feasible values of electrostatic fields. For instance, in the range of distances considered in this figure, we see that the changes from attraction to repulsion occur within the values of $0.4 - 1.1 \times 10^{5}$~V/m. Notice that increasing the distance $a$, the electrostatic field necessary to overcome the dispersive force decreases, as expected, since the dispersive contribution diminishes with $a$ faster than the electrostatic one. Figure \ref{fig: PltHvara} presents a profile that also follows the previous discussion, showing the ratio $\Gamma$ as a function of the distance $a$ for different field intensities and the same sphere's radius $R = 60$~nm and field orientation $\theta_0 = \pi/2$.

\begin{figure}[t!]
	\centering
	\includegraphics[scale =0.55]{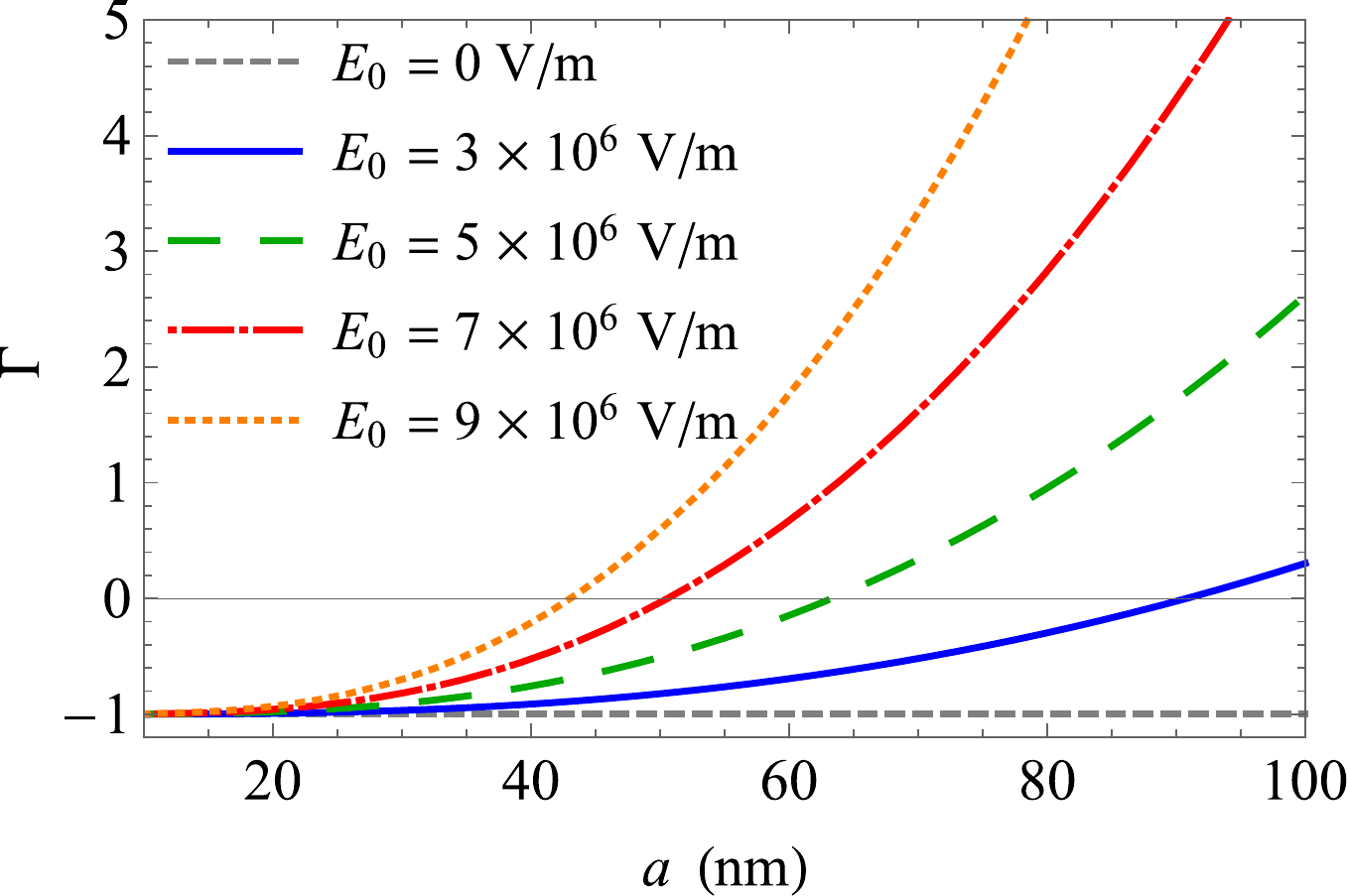}
	\caption{Ratio $\Gamma$ as a function of the distance $a$ from the atom to the sphere's surface for different field intensities in the short-distance regime. We set $R = 60$~nm and $\theta_0 = \pi/2$.}
	\label{curtasdistanciasouro}
\end{figure}

We call attention to the fact that $\Gamma$ depends only slightly on the sphere's radius. Indeed, in the small sphere limit ($R \ll a$), we may approximate the electrostatic force by Eq. (\ref{fzeldip}), as already discussed. A comparison with Eq. (\ref{UCPRed}) shows that the ratio $\Gamma$ is independent of $R$ for a given distance between the atom and the sphere's center (i.e., for fixed $z_a$). This is a remarkable feature as it holds regardless of the material composing the sphere and implies that the electrostatic control of the atom-sphere interaction is robust against the sphere's size, as long as the dipole approximation remains valid. However, when the atom-sphere distance is on the order of the sphere's radius or smaller, Eq. (\ref{UCPRed}) no longer holds. Assuming that this distance is short enough so that we can describe the dispersive interaction in the non-retarded limit, we must change our approach and calculate the ratio $\Gamma$, evaluating the dispersive contribution from Eq. (\ref{buhmanneq}), instead of from Eq. (\ref{UCPRed}). This situation is displayed in Fig. \ref{curtasdistanciasouro}, in which we plot $\Gamma$ as a function of shorter distances between the atom and the sphere's surface. Note that, for electric field intensities $30 - 100$ times the values depicted in Fig. \ref{fig: PltHvara}, we already overcome the dispersive force for distance regimes smaller than $100$~nm.

Analytical expressions both in the retarded and the non-retarded regimes can be obtained, providing us with interesting physical insights for our results. In this paragraph, we will not focus on hydrogen but rather on any atom described by the two-level model given in Eq. (\ref{alpha}). In the retarded regime ($z_a \gg c/\omega_0$), we may determine the dispersive force from Eq. (\ref{udr}) and employ Eq. (\ref{fzeldip}) with $\theta_0 = \pi/2$ to obtain 
\begin{equation}
	\Gamma  = \frac{48 \pi^2 \varepsilon_0 E_0^2 z_a^4}{161 \hbar c} - 1\, .
\end{equation}
This expression has some notable characteristics. It is independent of the atomic parameters and the material composing the sphere (but, to the latter, we must also assume that the retarded limit is true for frequencies that characterize the sphere's material). This is an expected result since, in the retarded regime, the dispersive force is proportional to the static polarizability that cancels out when we evaluate $\Gamma$. We see that the electrostatic force becomes relevant when the electrostatic energy $U^{\rm (el)} \sim \varepsilon_0 E_0^2 z_a^3$, corresponding to the electrostatic energy contained in a region of the size of the distance separating the atom and the sphere, is on the order of the energy carried by the photons relevant to the dispersive interaction, which is given by $U_{\rm photon} \sim \hbar c/z_a$. To give a numerical estimate, if one takes the distance to be ten times the hydrogen wavelength transition, it would be necessary an electric field intensity around $10^{-3}$~V/m to overcome the dispersive interaction. This small field reflects a feeble dispersive force. However, situations of more practical interest to the applications mentioned in Sec. \ref{SecIntrod} concern the short distance limit. In this non-retarded regime, when calculating the dispersive force from Eq. (\ref{udnr}) and using Eq. (\ref{fzeldip}) with $\theta_0 = \pi/2$, we arrive at
\begin{equation}
	\Gamma = \frac{4 \pi \varepsilon_0 E_0^2 z_a^3}{3 \hbar \omega_0} - 1 \, .
\end{equation}
Therefore, the minimum field $E_c$ necessary for the electrostatic attraction to overcome the dispersive force satisfies the relation
\begin{equation}
	\frac{4 \pi \varepsilon_0 E_c^2 z_a^3}{3} = \hbar \omega_0 \, , 
\label{campocriticolondon}
\end{equation}
which means that repulsion is ensured once the atomic transition energy equals the electrostatic energy contained in a sphere of radius corresponding to the atom-sphere distance. We remind the reader that the results obtained in this section did not assume the non-retarded regime nor a perfectly conducting sphere. Nonetheless, the power-law present in Eq. (\ref{campocriticolondon}) is valuable in order to provide a first estimate regarding the necessary electric field to achieve repulsion. For example, for a hydrogen atom and taking $z_a = 800$~nm, we obtain $E_c \sim 6 \times 10^4$~V/m, a value on the same order of magnitude and only $20\%$ above the predicted one using a more general expression, given in Eq. (\ref{UCPRed}). Expression (\ref{campocriticolondon}) overestimates the necessary field for two reasons: {\it (i)} the non-retarded limit overestimates the dispersive force and {\it (ii)} a perfectly conductor demands a stronger field than a real metal (see Sec. \ref{SubSecC}).

Another important aspect included in Eq. (\ref{campocriticolondon}) is that the required electrostatic field in the non-retarded regime scales with $\sqrt{\omega_0}$ and, consequently, atoms with lower transition frequencies are easier to control. It can be well understood if we notice that when we increase $\omega_0$, the dispersive and the electrostatic forces diminish since the atomic polarizability is reduced. Nevertheless, as can be inferred from Eqs. (\ref{UCPRed}), (\ref{fzelpisobre2}), and (\ref{alpha}), the integrand in the ratio $F^{\rm (disp)}/F_z^{\rm (el)}$ depends on $\omega_0$ through the factor $\omega_0^2/(\omega_0^2 + \xi^2)$. Therefore, increasing $\omega_0$ also increases the integrand for every value of $\xi$. As a consequence, as we increase $\omega_0$, the dispersive force weakens more slowly than the electrostatic one, thus explaining the aforementioned behavior. Physically, it is related to the fact that the dispersive interaction is a fluctuating-induced phenomenon that arises in the coupling between the dipole fluctuations on each body. Although increasing the transition frequency reduces the dispersive force (since it weakens the virtual excitations and reduces the polarizability), it also allows for a faster atomic dipole fluctuation, which enhances the dipole-dipole correlation and smoothes the decrease in the dispersive force in comparison with the electrostatic force. This hallmark was discussed for two-level atoms in the non-retarded regime but, in the following subsection, we demonstrate that it remains accurate even for a more realistic treatment of atoms and taking into account the complete dispersive energy interaction, dropping the non-retarded interaction assumption.


\subsection{Other atoms}


\begin{figure}[b!]
	\centering
	\includegraphics[scale =0.5]{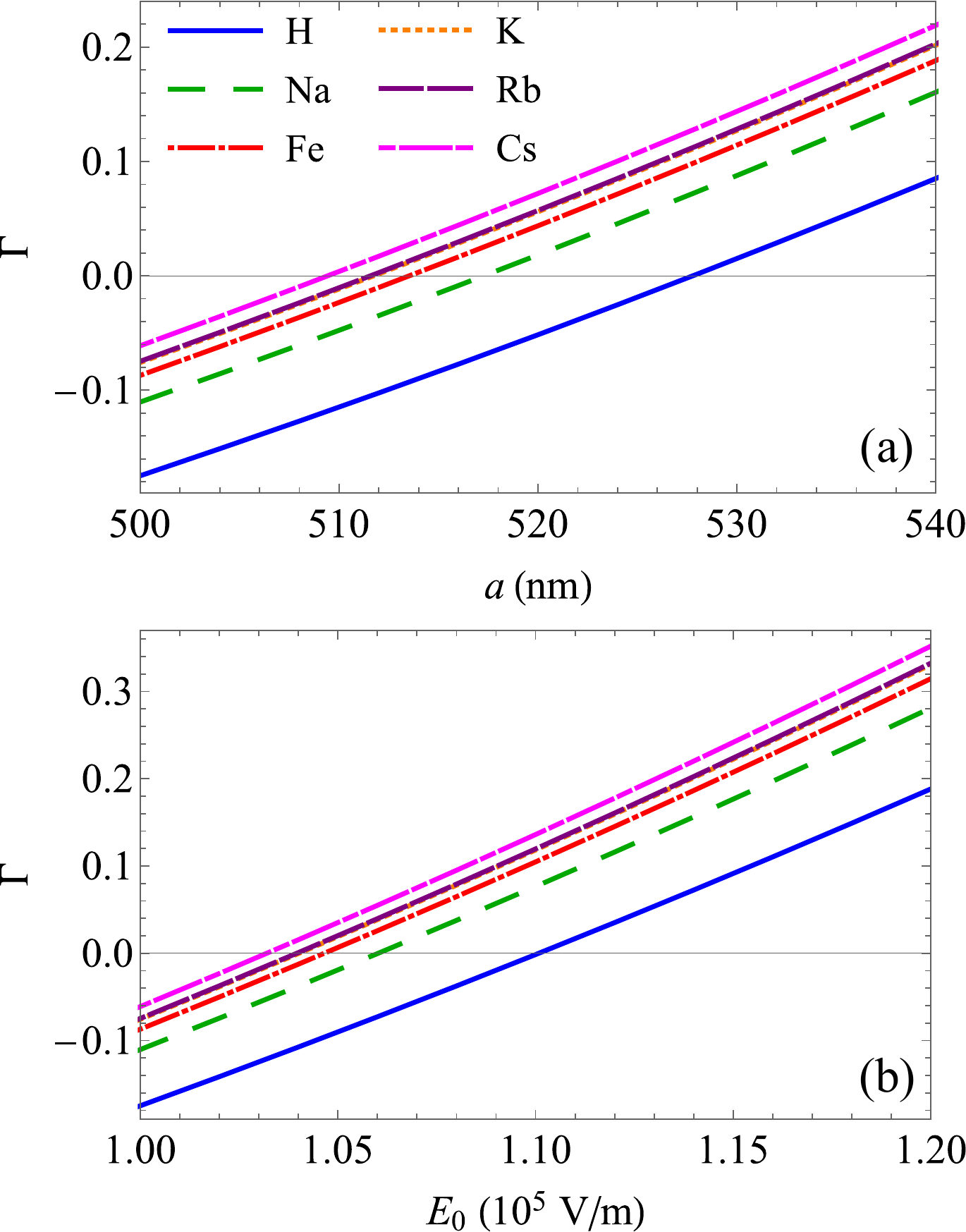}
	\caption{{\bf (a)} Ratio $\Gamma$ as a function of the distance $a$ between a given atomic species and the sphere's surface for $E_0 =1 \times 10^5$~V/m. {\bf (b)} Ratio $\Gamma$ as a function of the field intensity $E_0$ with $a = 500$~nm. In all panels, different colors refer to different atomic species and we set $R = 60$~nm and $\theta_0 = \pi/2$.}
	\label{fig: Pltvara}
\end{figure}

Even though the two-level approximation suits well in the mathematical description of the hydrogen polarizability, there is no good agreement with experiments when assuming only one relevant transition for heavier atoms. However, in such cases, a good solution is to implement the so-called two-oscillator model, which consists of keeping two atomic transitions as the main contributions to the dynamical atomic polarizability, as shown in the Appendix. We now come back to the description of the dispersive interaction through Eq. (\ref{UCPRed}) and, by using Eqs. (\ref{fzelpisobre2}) and (\ref{2oscillator}), we are guided to the results presented in Fig. \ref{fig: Pltvara}. The first panel exhibits $\Gamma$ as a function of the distance $a$ for a fixed field intensity $E_0 =1 \times 10^5$~V/m, while the second panel exposes the ratio $\Gamma$ as a function of the electric field at a given distance $a = 500$~nm from the sphere's surface. In both of them, each curve denotes a different atomic specimen. Note from Fig. \ref{fig: Pltvara}(a) that all $\Gamma$ curves eventually cross the zero value when $a$ ranges between $508 - 530$ nm. It should also be noticed that it is possible to change the profile of the net force on each atom just by varying its distance $a$. An equivalent route of observing this same behavior is to investigate $\Gamma$ as a function of the electric field and, accordingly, the outcome is the possibility of changing the $\Gamma$ sign just by varying the intensity of this external agent. Moreover, we stress that this modification occurs for feasible values of the electric field, being within the scope of experimental realization. The pattern presented in the curves of Figs. \ref{fig: Pltvara}(a) and \ref{fig: Pltvara}(b) can also be understood. In the two-oscillator model [Eq. (\ref{2oscillator})], one of the frequencies is generally much smaller than the other (typically, $\omega_{01}/\omega_{02}\sim 0.01$) and, as a consequence, the lower transition frequency ends up being responsible for the dominant response. Hence, the electrostatic effect is more prominent for atoms exhibiting smaller transition frequencies, exactly as illustrated in the panels. In other words, for a given distance, atoms with smaller $\omega_{01}$ requires a smaller electrostatic field to overcome the dispersive force.

\begin{figure}[t!]
	\centering
	\includegraphics[scale =0.5]{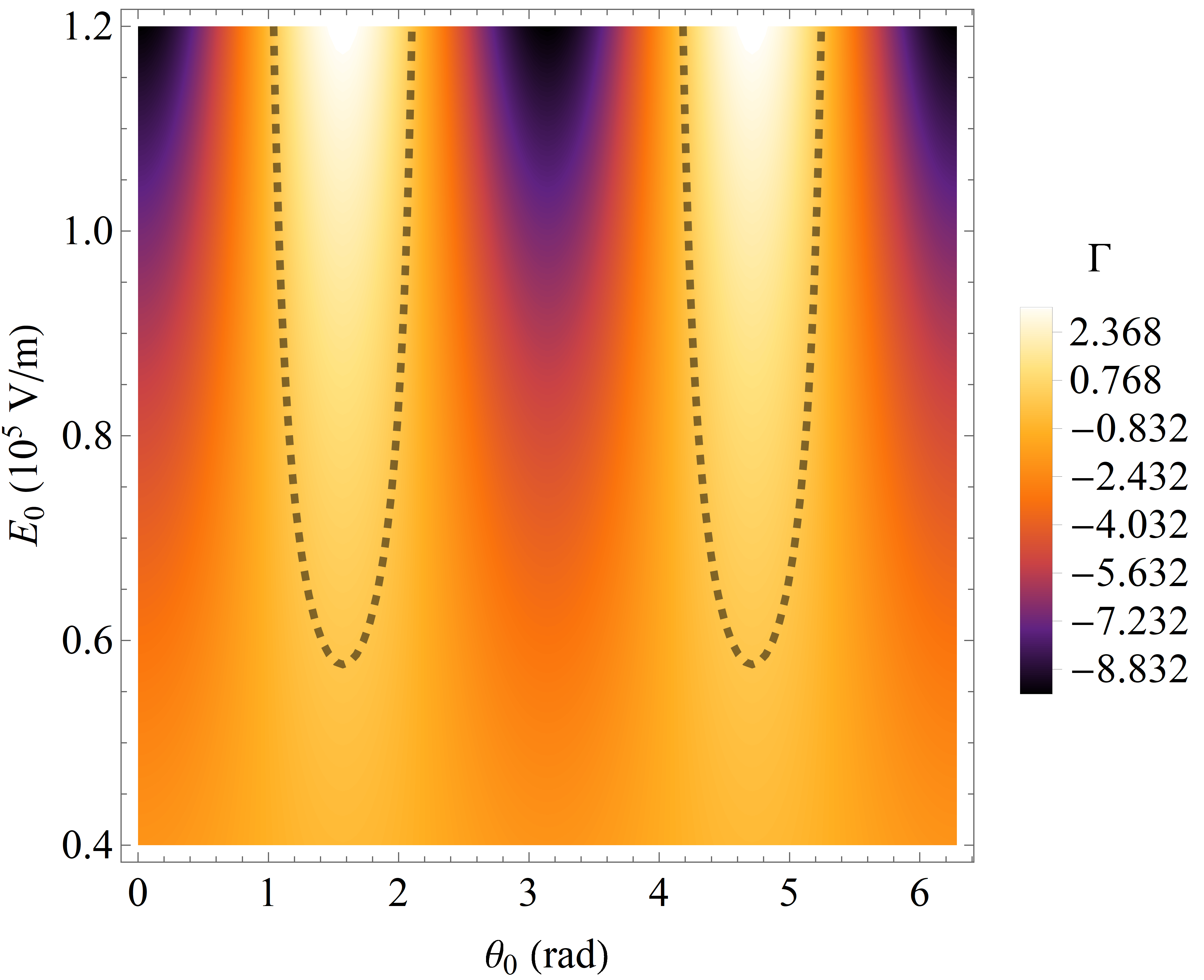}
	\caption{Contour plot of the ratio $\Gamma$ for the cesium atom varying $\theta_0$ and $E_0$. The dashed black line indicates combinations of $\theta_{0}$ and $E_{0}$ for which $\Gamma = 0$. Here, we chose $R = 60$~nm and $a = 700$~nm.}
	\label{fig: CscontourthE}
\end{figure}

\begin{figure}[t!]
	\centering
	\includegraphics[scale =0.5]{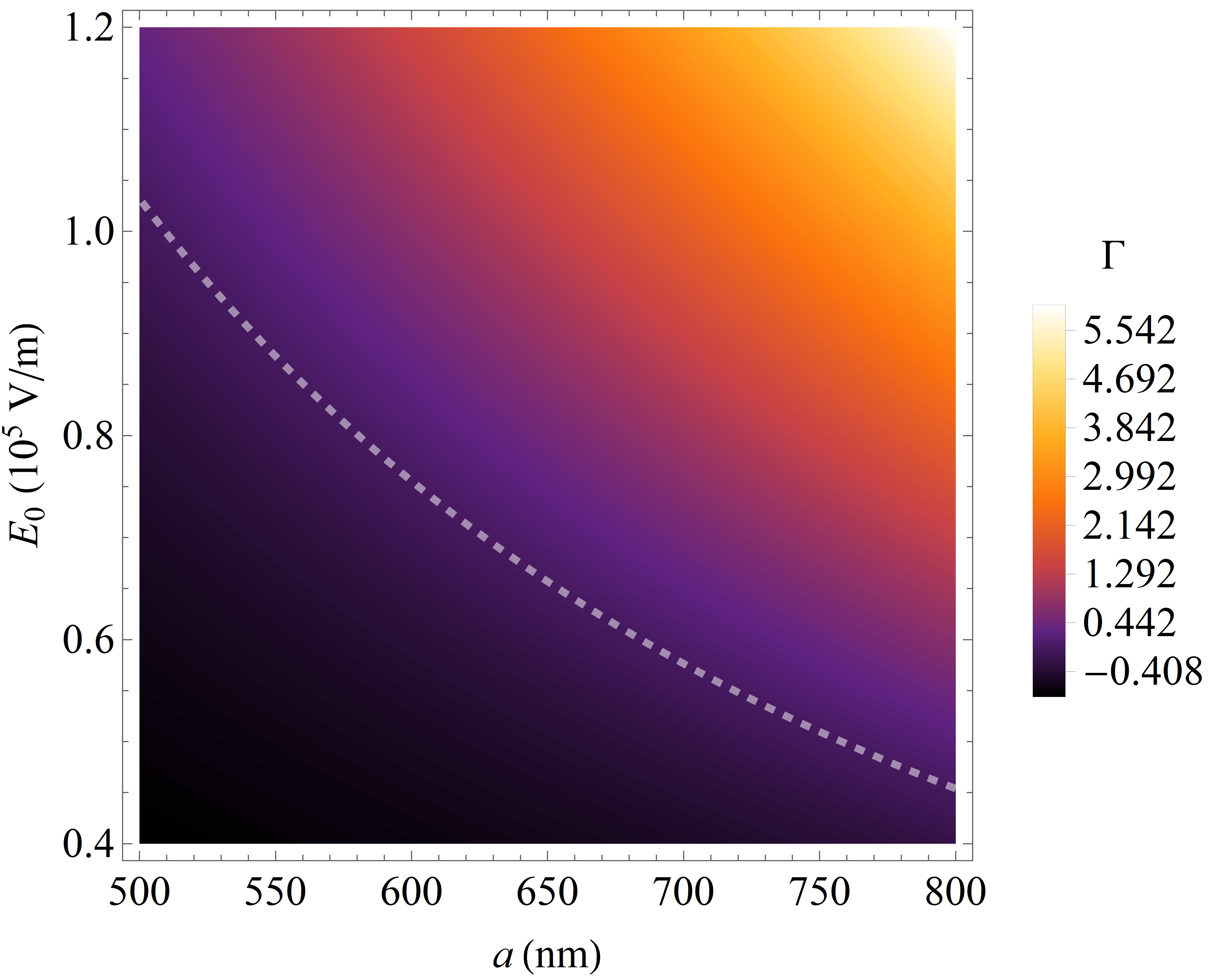}
	\caption{Contour plot of the ratio $\Gamma$ for the cesium atom varying $a$ and $E_0$. The dashed white line indicates combinations of $a$ and $E_{0}$ for which $\Gamma = 0$. Here, we chose $R = 60$~nm and $\theta_0 = \pi/2$.}
	\label{fig: CscontouraE}
\end{figure}

Amid the atoms we chose to work with, cesium is the most easily controlled by the electrostatic field, because it exhibits the smallest $\omega_{01}$ (although it has the highest $\omega_{02}$) -- except for hydrogen, which is well represented by the one-oscillator model, as we stated before. Therefore, we have chosen this atomic species to explore the contour plots that follow. We begin with Fig. \ref{fig: CscontourthE} that shows $\Gamma$ as a function of $\theta_0$ and $E_0$. In this figure, the dashed black line indicates $\Gamma = 0$, separating regions in the parameter space of $\theta_{0}$ and $E_{0}$ for which the net force is attractive or repulsive. Note that the most intense repulsion value (when it takes place) occurs for $\theta_{0} = \pi/2$, as previously mentioned. Curves similar to the ones in Fig. \ref{fig: PltHvarth700}, but to the cesium atom, are found by taking horizontal lines of constant $E_0$ in Fig. \ref{fig: CscontourthE}. Lastly, in Fig. \ref{fig: CscontouraE}, we consider $\Gamma$ as a function of $a$ and $E_0$, and, once more, the dashed white line represents $\Gamma = 0$. Likewise, by selecting the horizontal lines, we obtain results similar to those in Fig. \ref{fig: PltHvara}.

\subsection{Atoms near a dielectric sphere \label{SubSecC}}


\begin{figure}[b!]
	\centering
	\includegraphics[scale =0.5]{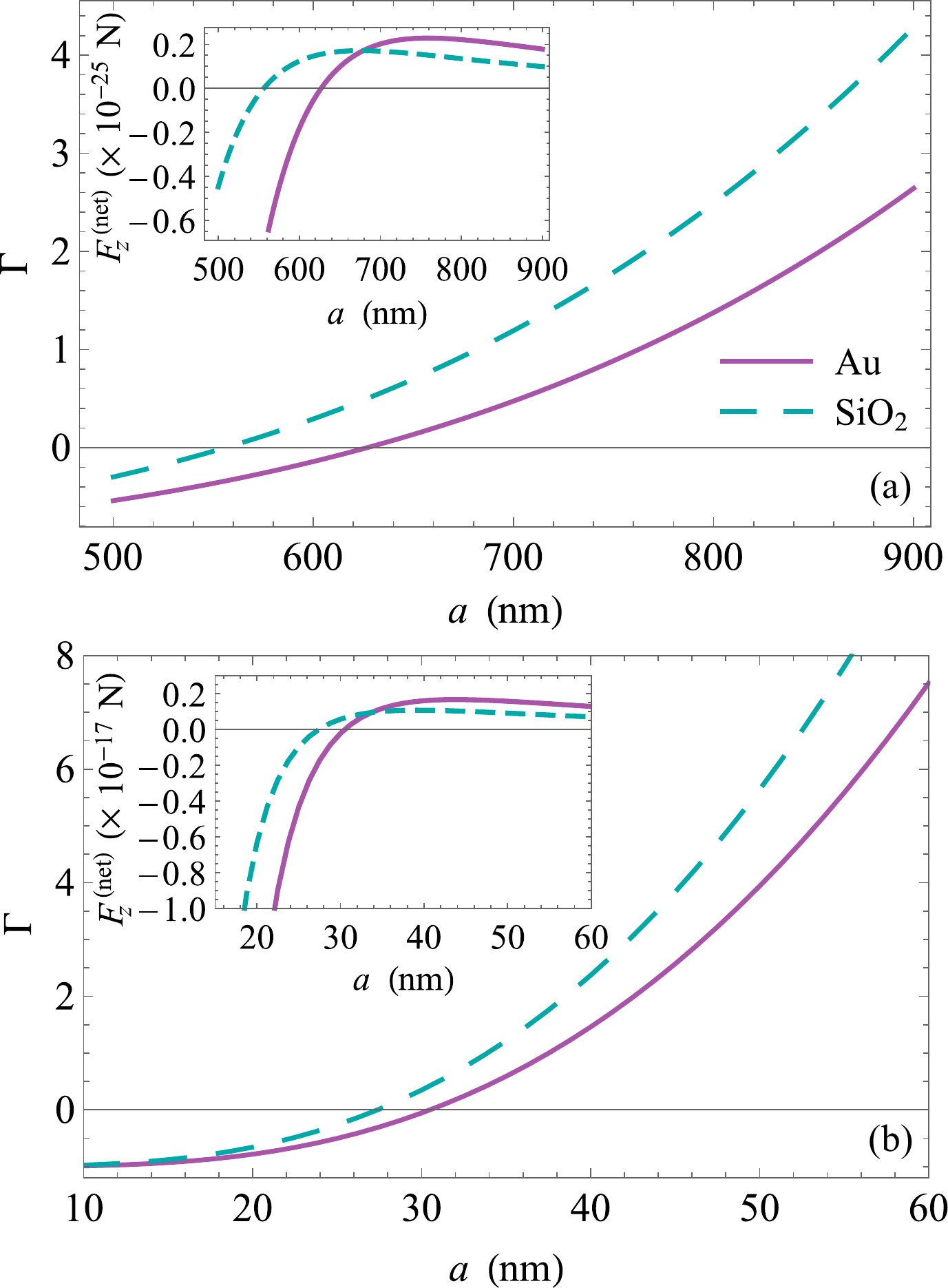}
	\caption{Ratio $\Gamma$ as a function of the distance $a$ between a cesium atom and the sphere's surface for {\bf (a)} $E_0 = 7 \times 10^4$~V/m, using Eq. (\ref{UCPRed}), and {\bf (b)} $E_0 = 9 \times 10^6$~V/m, using Eq. (\ref{buhmanneq}). In all panels, different colors refer to spheres made of different materials, and we set $R = 60$~nm and $\theta_0 = \pi/2$. The insets show the resultant force on the atom as a function of $a$.}
	\label{fig: MetDie}
\end{figure}

In our previous discussions, we have always assumed the interaction with a metallic sphere. If the atoms are near a dielectric one instead, the general expressions presented in Sec. \ref{AtomSphereInt} are still valid and we must only substitute back on them the polarizability of the dielectric sphere in question. A model for the electric permittivity of silica spheres is described in the Appendix. For dielectrics, both the dispersive and the electrostatic interactions are weaker than for the metallic case, when considering a given atom placed at a fixed distance separating it from the sphere. Therefore, it is not clear at first sight whether $\Gamma$ should be greater for a dielectric or a metallic sphere (with all conditions being the same). In other words, is it easier to generate repulsion when an atom is close to a dielectric instead of a metallic sphere? To address this question, in Fig. \ref{fig: MetDie}, we plot $\Gamma$ for a cesium atom next to a silica sphere as a function of the atom-sphere distance $a$ in different regimes and compare it with $\Gamma$ for the same situation replacing the silica sphere for a gold one with the same radius. We see that, for a fixed distance, $\Gamma$ is always greater in the dielectric case. For example, the distance for which the electrostatic force equalizes the dispersive one for the gold case is a distance where the same field has already generated repulsion in the dielectric situation. It is related to the results obtained in Sec. \ref{SubSecA}, where we explained the mechanism which makes it easier to control atoms with smaller transition frequencies. Metals allow for a faster response than dielectrics (analogous to atoms with higher transition frequencies), implying greater polarizability for each imaginary frequency, which, in turn, enhances the dispersive and the electrostatic forces. However, the effect on the former is more pronounced due to its fluctuating-induced nature, and, therefore, metals require stronger electric field intensities than dielectrics so that the atom-sphere force becomes repulsive.


\section{Final remarks and conclusions}
\label{Conclusions}


We have investigated the possibility of controlling the dispersive interaction in the system composed of an atom and a neutral and isolated sphere when exposed to an external electrostatic field. We have explored both metallic and dielectric spheres as well as the implications of considering different atomic species. Our results demonstrated that the electrostatic force, that arises between the atom and the sphere, can provide active control of this interaction without demanding physical contact. More specifically, the electrostatic force enables the tunability of the sign of the resultant force since this electrostatic contribution can overcome the dispersive one. Moreover, we highlight that such a degree of control can be achieved for feasible values of the electric field, being within the scope of experimental realization. Concerning the outcomes of studying different atomic species and materials composing the sphere, we concluded that larger field intensities are required when dealing with metallic spheres and with atoms that exhibit larger transition frequencies. We have also discussed in detail the dependence of our results on the magnitude of the electric field, its orientation, and the atom-sphere distance. Although our results assume a spherical surface, we expect that the orders of magnitude of the electrostatic field required to overcome the dispersive interaction are insensitive to small shape variations. Nonetheless, it is still interesting to be able to investigate different geometries. Whenever the dipole approximation remains valid, our treatment can be immediately applied. Therefore, it suffices to substitute the polarizability of the sphere with the appropriate dynamical polarizability of the object under study. As the polarizability scales with the volume, the orders of magnitude involved must hold regardless of the shape, replacing the object with a sphere of the same volume. We expect that the results presented above may open different routes to control dispersive forces and inspire other configurations that may also exhibit repulsion.


\begin{acknowledgments}

We are indebted to S. Y. Buhmann for valuable suggestions and for calling our attention to Ref. \cite{Hemmerich2016} and to P. A. Maia Neto for enlightening discussions. We also thank the Brazilian agencies FAPERJ, CAPES, and CNPq for financial support (C.F. was supported under Grant No. CNPq, 310365/2018-0).

\end{acknowledgments}


\appendix


\section{Atomic polarizabilities and dielectric functions \label{AppendA}}


In order to characterize the hydrogen atomic polarizability in the dispersive interaction, we employed a single-oscillator model, given by
\begin{equation}
	\alpha_a (i \xi) = \frac{\omega_0^2 \alpha_a (0)}{\omega_0^2 + \xi^2} \,,
\label{alpha}
\end{equation}
with $\alpha_a (0) = 4.5$~a.u. and $\omega_0 = 11.65$~eV \cite{woods}. In the case of heavier atoms, such as Na, K, Fe, Rb, and Cs, we employed a two-oscillator model, written as
\begin{equation}
	\alpha_a (i \xi) = \frac{\omega_{01}^2 \alpha_{a1} (0)}{\omega_{01}^2 + \xi^2} + \frac{\omega_{02}^2 \alpha_{a2} (0)}{\omega_{02}^2 + \xi^2} \,.
\label{2oscillator}
\end{equation}

\noindent The fitted parameters for each of these atomic specimen also analyzed here are reported in Table \ref{table1}.

\begin{table}[h!]

	\centering
	\begin{tabular}{||c c c c c c c c c||} 
		\hline
		Atom & & $\alpha_{a1} (0)$ (a.u.) & &  $\omega_{01}$ (eV) & & $\alpha_{a2} (0)$ (a.u.) & &  $\omega_{02}$ (eV) \\ [0.3ex] 
 		\hline
		\hline
		Na & & 162.1 & & 2.12 & & 0.547 & & 116.4 \\ [0.3ex] 
		K & & 288.4 & & 1.66 & & 1.754 & & 87.0 \\ [0.3ex] 
		Fe & & 307.8 & & 1.75 & & 9.972 & & 42.8 \\ [0.3ex] 
		Rb & & 316.7 & & 1.65 & & 1.85 & & 119.6 \\ [0.3ex] 
		Cs & & 397.3 & & 1.53 & & 2.597 & & 123.8 \\ [0.3ex] 
		\hline
	\end{tabular}
	\caption{Data for Na, K, Fe, Rb, and Cs  atoms. This table contains parameters of the two-oscillator model to be used in Eq. (\ref{2oscillator}) \cite{woods} ($1$ a.u. $= 1.648 \times 10^{-41}$ C$^2$m$^2$J$^{-1}$).}
	\label{table1}

\end{table}

\begin{table}[h!]

	\centering
	\begin{tabular}{||c c c c c c c||} 
		\hline
		Parameter & & Value (Hz) & & Parameter & & Value (Hz) \\ [0.3ex] 
 		\hline
		\hline
		$\omega_{p0}$ & & $1.37 \times 10^{16}$ & & $\gamma_{0}$ & & $4.05 \times 10^{13}$ \\ [0.3ex] 
		$\omega_{p1}$ & & $1.75 \times 10^{14}$ & & $\gamma_{1}$ & & $4.28 \times 10^{13}$ \\ [0.3ex] 
		$\omega_{p2}$ & & $2.96 \times 10^{16}$ & & $\gamma_{2}$ & & $8.09 \times 10^{15}$ \\ [0.3ex] 
		$\omega_{T1}$ & & $1.32 \times 10^{14}$ & & $\omega_{T2}$ & & $2.72 \times 10^{16}$ \\ [0.3ex]
		\hline
	\end{tabular}
	\caption{Data for gold and silicon dioxide spheres. This table contains parameters for each model in Eqs. (\ref{dielfuncAu}) \cite{Ordal1985} and (\ref{dielfuncSiO2}) \cite{Hemmerich2016}.}
	\label{table2}

\end{table}

The mathematical description of the different materials composing the spheres is based on their dielectric functions. We assume that the expressions for the metallic (gold -- Au) and dielectric (silicon dioxide -- SiO$_2$) spheres are written using a Drude model and a Drude-Lorentz model, respectively, according to
\begin{align}
	\varepsilon_{\rm Au} (i \xi) &= 1 + \frac{\omega_{p0}^2}{\gamma_{0} \xi + \xi^2} \,, 
\label{dielfuncAu}\\
	\varepsilon_{\textrm{SiO$_2$}} (i \xi) &= 1 + \frac{\omega_{p1}^2}{\omega_{T1}^2 + \gamma_1 \xi + \xi^2} + \frac{\omega_{p2}^2}{\omega_{T2}^2 + \gamma_2 \xi + \xi^2}\,.
\label{dielfuncSiO2}
\end{align}

\noindent The fitted parameters for these materials are reported in Table \ref{table2}.



\end{document}